\documentclass[12pt]{article}
\usepackage{epsfig}
\usepackage{amsmath,amssymb,amsbsy}
\usepackage{amsfonts}
\usepackage{bm}

\def\Journal#1#2#3#4{{#1} {\bf #2}, #3 (#4)}

\def\NPB{{\em Nucl. Phys.} B}

\def\PRL{\em Phys. Rev. Lett.}
\def\PRD{{\em Phys. Rev.} D}
\def\ZPC{{\em Z. Phys.} C}
\def\EPJC{{\em Eur. Phys. J.} C}

\def\be{\begin{equation}}
\def\ee{\end{equation}}
\def\bea{\begin{eqnarray}}
\def\eea{\end{eqnarray}}

\begin{document}
\begin{center}
{\Large\bf{Significant issues related to elastic scattering\\
\vskip 0.3cm
at very high ensergies \footnote{Invited talk presented at the 14th Workshop on Elastic and Diffractive Scattering (EDS Blois Workshop), December 15-21, 2011, Qui Nhon, Vietnam}}}
\vskip1.4cm
{\bf Jacques Soffer}
\vskip 0.2cm
{\it Physics Department, Temple University\\
Barton Hall, 1900 N, 13th Street\\
Philadelphia, PA 19122-6082, USA}
\vskip 0.5cm
\end{center}

\begin{center}
{\bf Abstract}\\
\end{center}
After giving a short review on the impact picture approach for the elastic scattering amplitude, we 
will discuss the importance of some issues related to its real and imaginary parts. This will be illustrated in the
context of recent data from RHIC, Tevatron and LHC.

\section{Introduction}
The measurements of high energy  $\bar p p~\mbox{and}~p p$ elastic at ISR, SPS, and Tevatron colliders
have provided usefull informations on the behavior of the scattering
amplitude, in particular, on the nature of the Pomeron. A large step in energy
domain is accomplished with the LHC collider presently running, giving a unique
opportunity to improve our knowledge on the asymptotic regime of the scattering
amplitude and to verify the validity of our approach. We will first
recall the basic ingredients of the BSW amplitude and its essential features. We
will also mention the success of its predictions so far in the energy range below the LHC energy, for the total cross section
$\sigma_{tot}(s)$, the ratio of the real to imaginary parts of the forward amplitude $\rho(s)$ and the differential cross section 
${d\sigma (s,t)/dt}$. Our predictions at LHC will be shown
and compared with the first experimental results and we will recall why its is so important to measure $\rho$ at LHC 

\section{The BSW model}
The BSW model was first proposed, in 1978 \cite{bsw}, to describe the experimental data on elastic $p p$ and $\bar{p} p$, taken at the relatively low energies available
to experiments, forty years ago or so. Some more complete analysis were done later \cite{bsw1,bsw2,bsw3},  showing very successful theoretical predictions for these processes. Since a new energy domain is now accessible with the LHC collider at CERN \cite{deile}, it is a good time to recall the main features of the BSW model and to check its validity. The spin-independent elastic scattering amplitude is given by
\begin{equation}
  \label{eq:one}
  a(s,t) = \frac{is}{2\pi}\int e^{-i\mathbf{q}\cdot\mathbf{b}} (1 - 
e^{-\Omega_0(s,\mathbf{b})})  d\mathbf{b} \ ,
\end{equation}
where $\mathbf q$ is the momentum transfer ($t={-\bf q}^2$) and 
$\Omega_0(s,\mathbf{b})$ is the opaqueness at impact parameter 
$\mathbf b$ and at a given energy $s$, the square of the center-of-mass energy. We take the simple form
\begin{equation}
\label{eq:omega}
\Omega_0(s,\mathbf{b}) = S_0(s)F(\mathbf{b}^2)+ R_0(s,\mathbf{b}) \, ,
\end{equation}
the first term is associated with the "Pomeron" exchange, which generates 
the diffractive component of the scattering and the second term is 
the Regge background which is negligible at high energy.
The function $S_0(s)$ is given by the complex symmetric expression, obtained from the high energy behavior
of quantum field theory \cite{chengWu}
\begin{equation}
  \label{eq:Sdef}
  S_0(s)= \frac{s^c}{(\ln s)^{c'}} + \frac{u^c}{(\ln u)^{c'}}~,
\end{equation}
with $s$ and $u$ in units of $\mbox{GeV}^2$, where $u$ is the
third Mandelstam variable. In Eq. (\ref{eq:Sdef}), $c$ and $c'$ are two 
dimensionless constants given above \footnote{In the Abelian case one finds $c' = 3/2$ and it was
conjectured that in Yang-Mills non-Abelian gauge theory
one would get $c' = 3/4$ (T.T. Wu private communication).}
in Table 1. That they
are constants implies that the Pomeron is a fixed Regge cut rather than a
Regge pole. For the asymptotic behavior at high energy and modest momentum
transfers, we have to a good approximation
\begin{equation}
  \label{eq:uAp}
  \ln u = \ln s -i\pi~,
\end{equation}
so that
\begin{equation}
  \label{eq:S2}
    S_0(s)= \frac{s^c}{(\ln s)^{c'}} + 
\frac{s^ce^{-i\pi c}}{(\ln s-i\pi)^{c'}}~.
\end{equation}
The choice one makes for $F(\mathbf{b}^2)$ is essential and we take the Bessel transform of
\begin{equation}
  \label{eq:FtildDef}
  \tilde{F}(t) = f[G(t)]^2\frac{a^2+t}{a^2-t}~,
\end{equation}
where $G(t)$ stands for the proton "` nuclear form factor"', parametrized similarly to the electromagnetic
form factor, with two poles
\begin{equation}
  \label{eq:Gdef}
  G(t)=\frac{1}{(1- t/m_1^2)(1- t /m_2^2)}~.
\end{equation}
The remaining four parameters of the model, $f$, $a$, $m_1$ $\mbox{and}$ $m_2$, are given in Table 1.\\
We define the ratio of the real to imaginary parts of the forward amplitude
\begin{equation}\label{eq:rho}
\rho(s) = \frac{\mbox{Re}~a(s, t=0)}{\mbox{Im}~a(s, t=0)} \,, 
\end{equation}
 the total cross section
\begin{equation}\label{eq:sigtot}
\sigma_{tot} (s) = \frac{4\pi}{s}\mbox{Im}~a(s, t=0) \,,
\end{equation}
the differential cross section
\begin{equation}\label{eq:dsigdt}
\frac{d\sigma (s,t)}{dt} = \frac{\pi}{s^2}|a(s,t)|^2 \,,
\end{equation} 
and the integrated elastic cross section
\begin{equation}\label{eq:sigel}
\sigma_{el} (s) = \int dt \frac{d\sigma (s,t)} { dt}\,.
\end{equation}

 \begin{table}
    \centering
        \caption{\label{tab:table1} Parameters of the BSW model \cite{bsw3}.}
        \vspace{0.4cm}
    \begin{tabular}{|rllrll|}\hline
{$c$} & = & 0.167, & \;\;{$c'$}&=&0.748\\
{$m_1$}&=&0.577 GeV,&\;\;{$m_2$}&=&1.719 GeV\\
{$a$}&=&1.858 GeV,&\;\;{${\it f}$}&=&6.971 GeV{$^{-2}$}\\\hline
    \end{tabular}
      \end{table}
      
\begin{figure}[htb]
\begin{center}
\psfig{figure=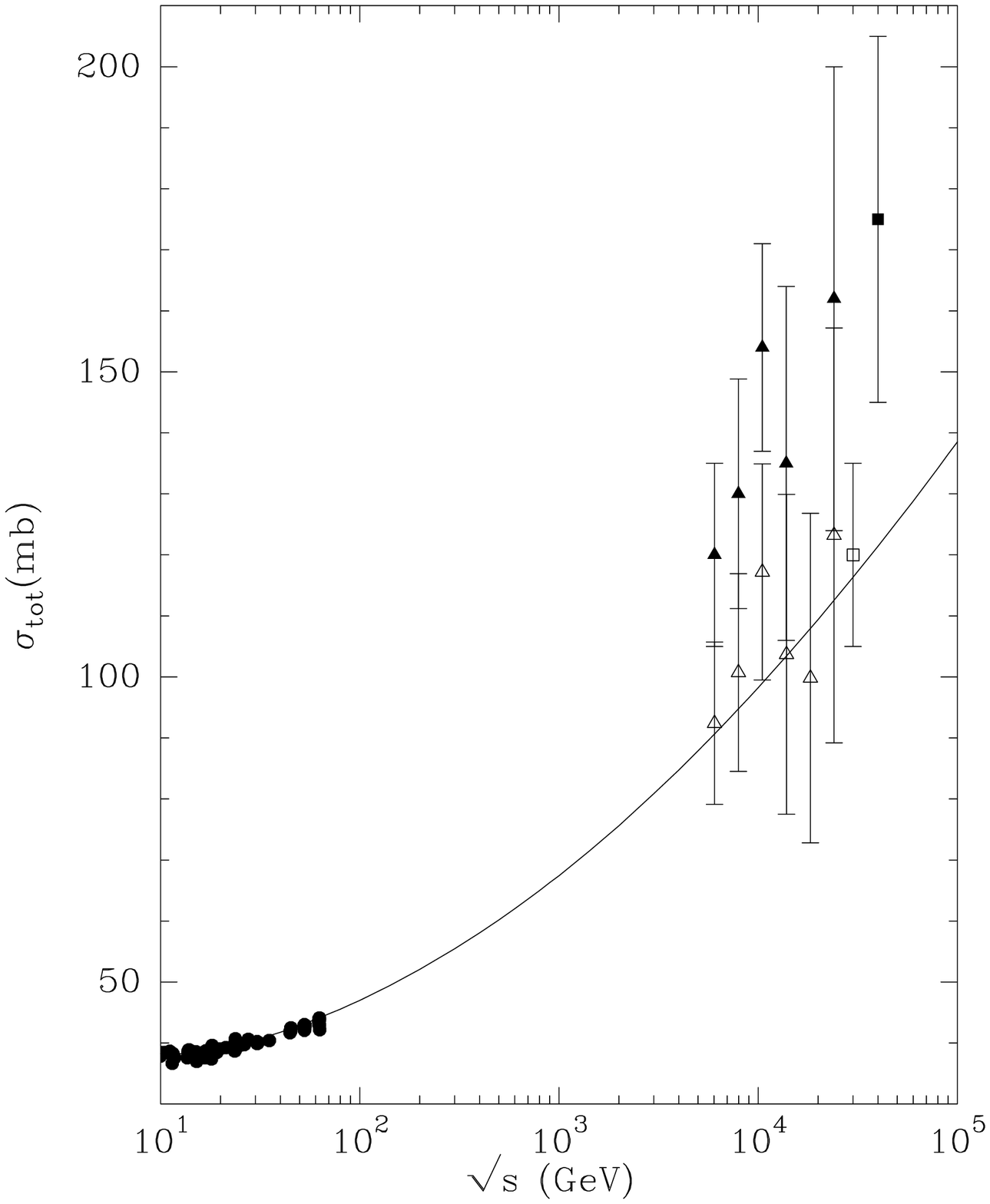,height=3.2in}
\psfig{figure=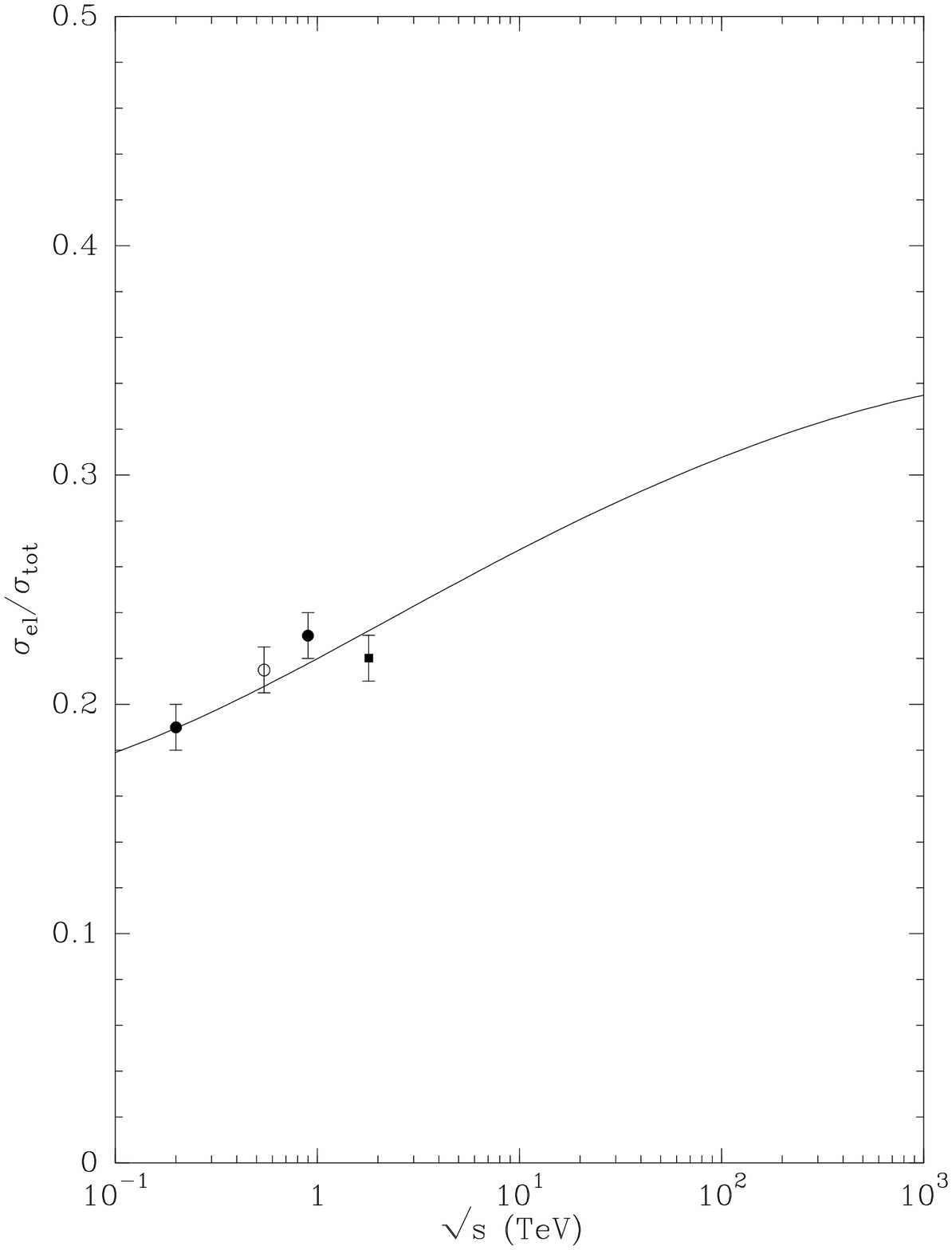,height=3.2in}
\end{center}
\caption{ $pp$ ($\bar pp$) elastic scattering, $\sigma_{tot}$, ({\it Left}), $\sigma_{el}/\sigma_{tot}$ ({\it Right}) as a function of the energy. (Taken from Ref.(4)).}
\label{fig:sig-totel}
\end{figure}
\section{Issues with the real and imaginary parts of the amplitude}
One important feature of the BSW model is, as a consequence of Eq. (\ref{eq:S2}), the fact that the phase
of the amplitude is built in. Therefore real and imaginary parts of the amplitude cannot be chosen independently and we will
now see how to test them, according to different $t$ regions.
\subsection{Forward region}
\begin{figure}[htb]
\begin{center}
  \begin{minipage}{6.5cm}
\includegraphics[width=6.8cm,height=7.0cm]{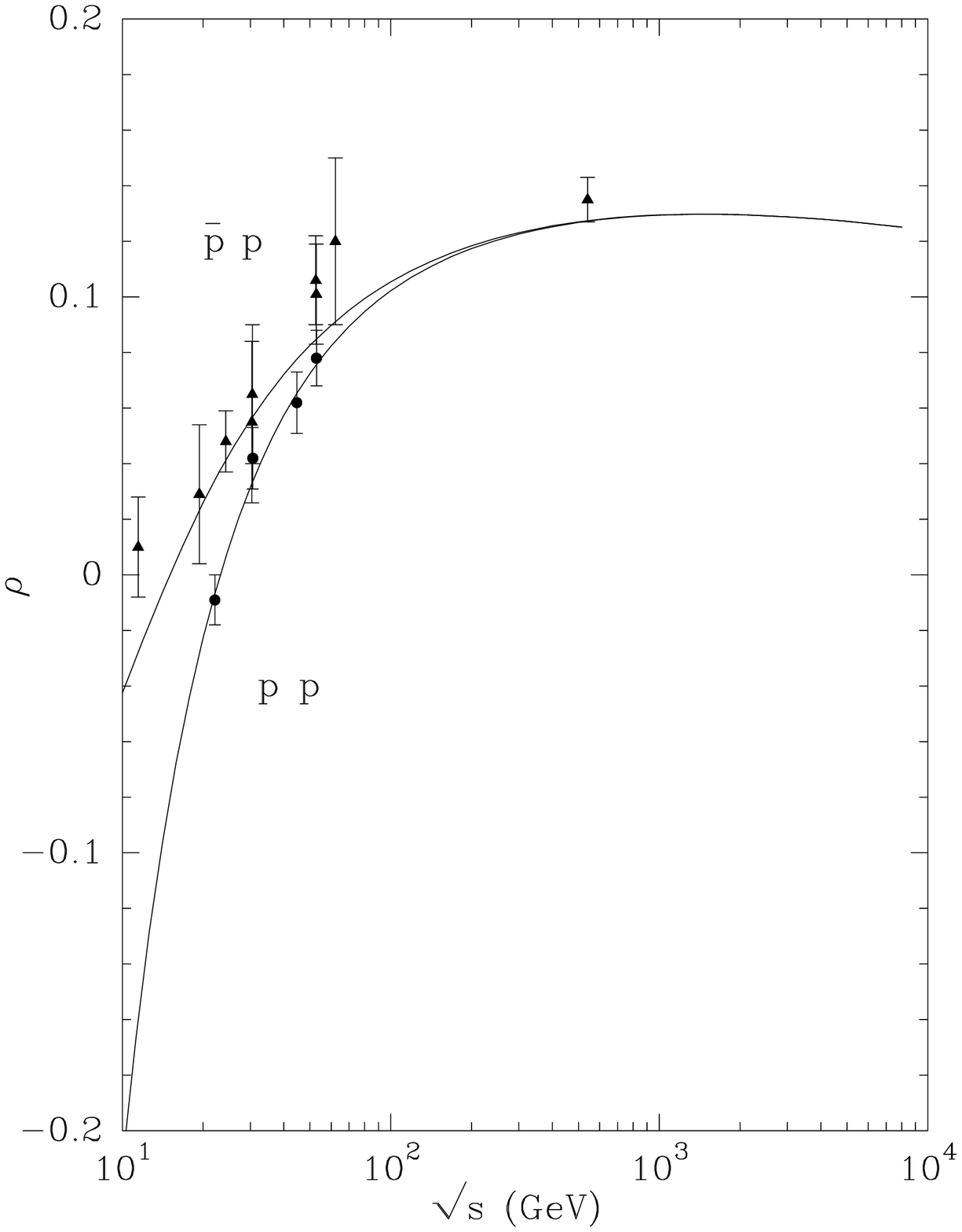}
  \end{minipage}
  \begin{minipage}{6.5cm}
\includegraphics[width=6.0cm,height=9.5cm]{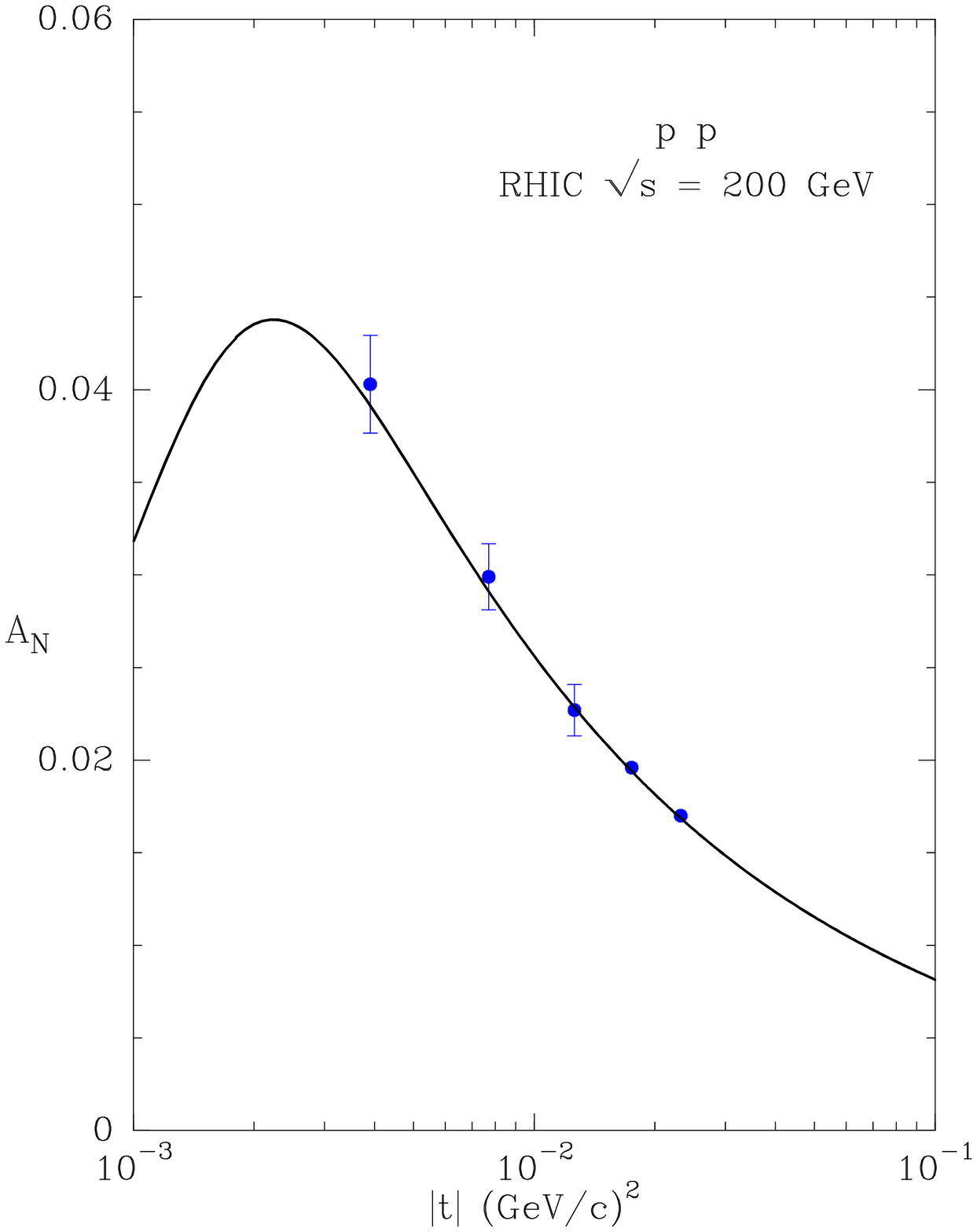}
  \end{minipage}
\end{center}
\vspace*{-10mm}
\caption{({\it Left}): $\rho$ for $pp$ ($\bar pp$) elastic scattering as a 
function of $\sqrt{s}$ (Taken from Ref. (4)). ({\it Right}): The analyzing 
power $A_N$ versus $t$ at RHIC energy. The prediction is taken from Ref. (8) 
and the data from Ref. (9).}
\label{fig:rho-asy}
\end{figure}

Consider first the total cross section which is directly related to $\mbox{Im}~a(s, t=0)$. We
show in Fig. \ref{fig:sig-totel}
 (Left) our prediction up to cosmic rays energy.
 The BSW approach predicts at 7 TeV $\sigma_{tot}= 93.6\pm 1 \mbox{mb}$. Two other important quantities are the integrated elastic cross section $\sigma_{el}$, which is predicted to be $\sigma_{el}= 24.8 \pm 0.3\mbox{mb}$ and finally the total inelastic cross section defined as $\sigma_{inel}=\sigma_{tot}-\sigma_{el}$.\\
These predictions must be compared with different new experimental LHC results \cite{deile}, namely, from TOTEM, 
$\sigma_{tot}= 98.3 \pm 0.2(stat) \pm 2.7(syst)\mbox{mb}$, $\sigma_{el}= 24.8 \pm 0.2(stat) \pm 1.2(syst)\mbox{mb}$ and
$\sigma_{inel}= 73.5 \pm 0.6(stat) + 1.8(-1.3)(syst) \mbox{mb}$, from ATLAS which has found $\sigma_{inel}= 69.4 \pm 2.4(expt) \pm 6.9(extra) \mbox{mb}$ and from CMS, which has reported $\sigma_{inel}= 68 \pm 2(syst) \pm 2.4(lum) \pm 4(extra) \mbox{mb}$.\\
We notice that our $\sigma_{inel}$ is in excellent agreement with the last two determinations, but although our $\sigma_{el}$ agrees
very well with the value of TOTEM, our prediction for $\sigma_{tot}$ is higher but consistent with their value.\\
Another specific feature of the BSW model is the fact that 
it incorporates the theory of expanding protons \cite{chengWu}, with the 
physical consequence that the ratio $\sigma_{el}/\sigma_{tot}$ 
increases with energy. This is precisely in agreement with the data, as
shown in Fig. \ref{fig:sig-totel} (Right), and when $s \to \infty$ one 
expects $\sigma_{el}/\sigma_{tot}\to 1/2$, which is the black disk limit.\\
The behavior of $\rho(s)$ with the energy is displayed in 
Fig. \ref{fig:rho-asy} (Left) and shows that the BSW model 
predicts the correct real part of the forward elastic amplitude. 
$\rho(s)$ appears to have a flat energy dependence in the high 
energy region and in the black disk limit $s \to \infty$ one 
expects $\rho(s)\to 0$ and at this stage it is very legitimate 
to ask the following question:\\
{\bf{Why should $\rho$ be measured at the LHC \cite{bkmst} ?}}
\begin{itemize}
\item Real and imaginary parts of the scattering amplitude must obey 
dispersion relations according to local quantum field theory
\item In string theory extra dimensions could generate observable non-local 
effects and therefore a violation of dispersion relations
\item We can make a simple model to break polynomial boundness in some 
regions of 
the analyticity domain, leading for example to $\rho = 0.21$ at 14TeV
\item According to the BSW model, which satisfies dispersion relations, one 
should find instead $\rho =0.122\pm0.003$
\item Dispersion relations could be also violated if $\sigma_{tot}$ beyond 
the LHC energy, behaves very differently, due to some new physics.
\item The highest energy where one has a reliable value of $\rho$ is at 
$\sqrt{s}=546\mbox{GeV}$, $\rho =0.135 \pm 0.007$, since the Tevatron 
value $\rho =0.140 \pm 0.069$ is useless
\end{itemize}
 For all these reasons one needs an accurate value of $\rho$ at LHC.\\
 Before moving to the non-forward region it is worth mentioning another test of the BSW amplitude, by means of the analyzing power $A_N$, near the very forward direction. In this kinematic region, the so called Coulomb nuclear interference (CNI) region, $A_N$ results from the interference of the Coulomb amplitude which is purely real, with the imaginary part of the hadronic non-flip amplitude, namely $a(s, t)$, if one assumes that there is no contribution from the single-flip hadronic amplitude \cite{bsw4}. This is what we have done in the calculation of the curve displayed in Fig. \ref{fig:rho-asy} (Right) compared to some new data from STAR which confirms the absence of single-flip hadronic amplitude and the right determination of $\mbox{Im}~a(s, t)$ in the CNI region \cite{star}.
\subsection{Non-forward region}
\begin{figure}[htb]
\begin{center}
\psfig{figure=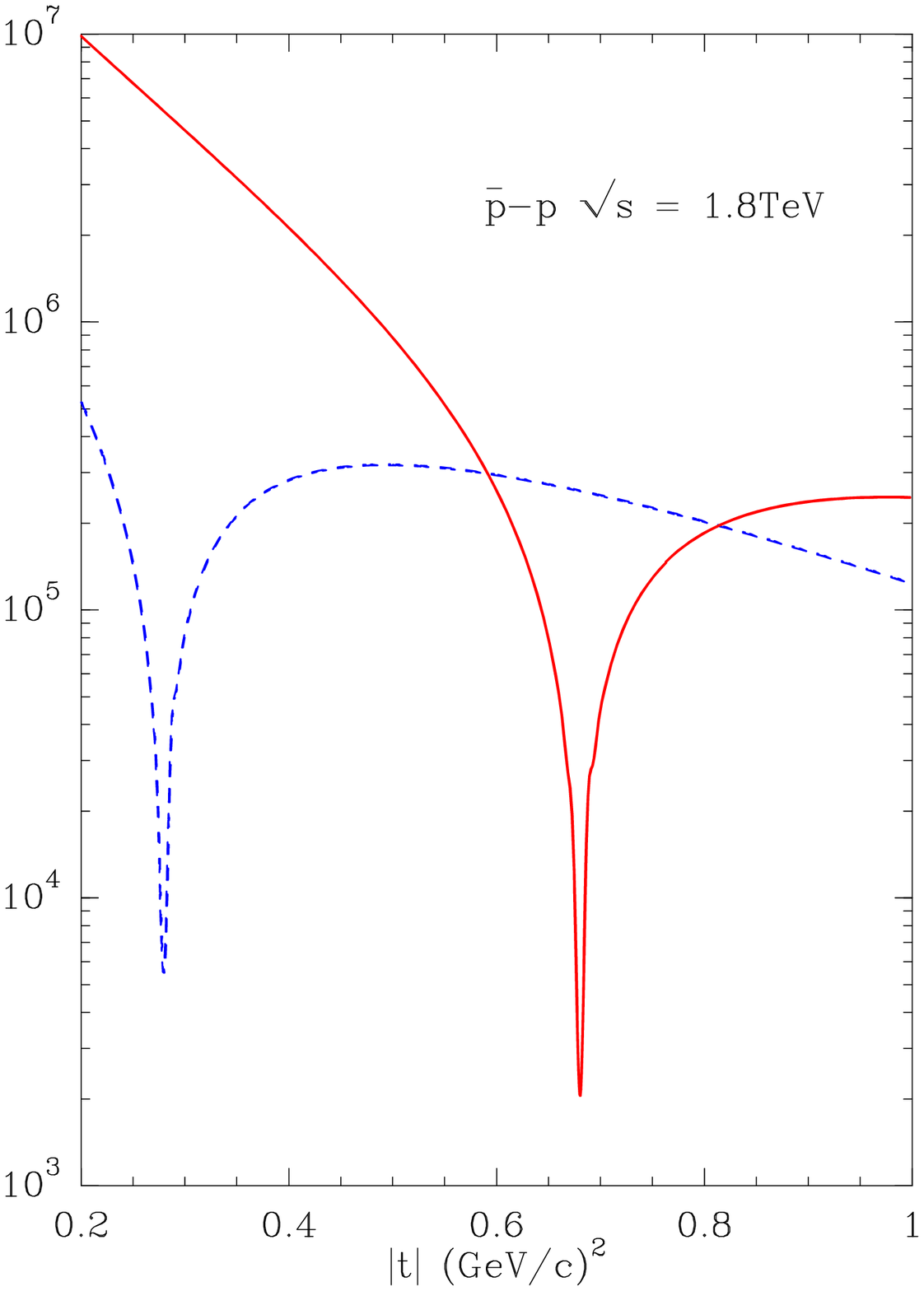,height=3.7in}
\psfig{figure=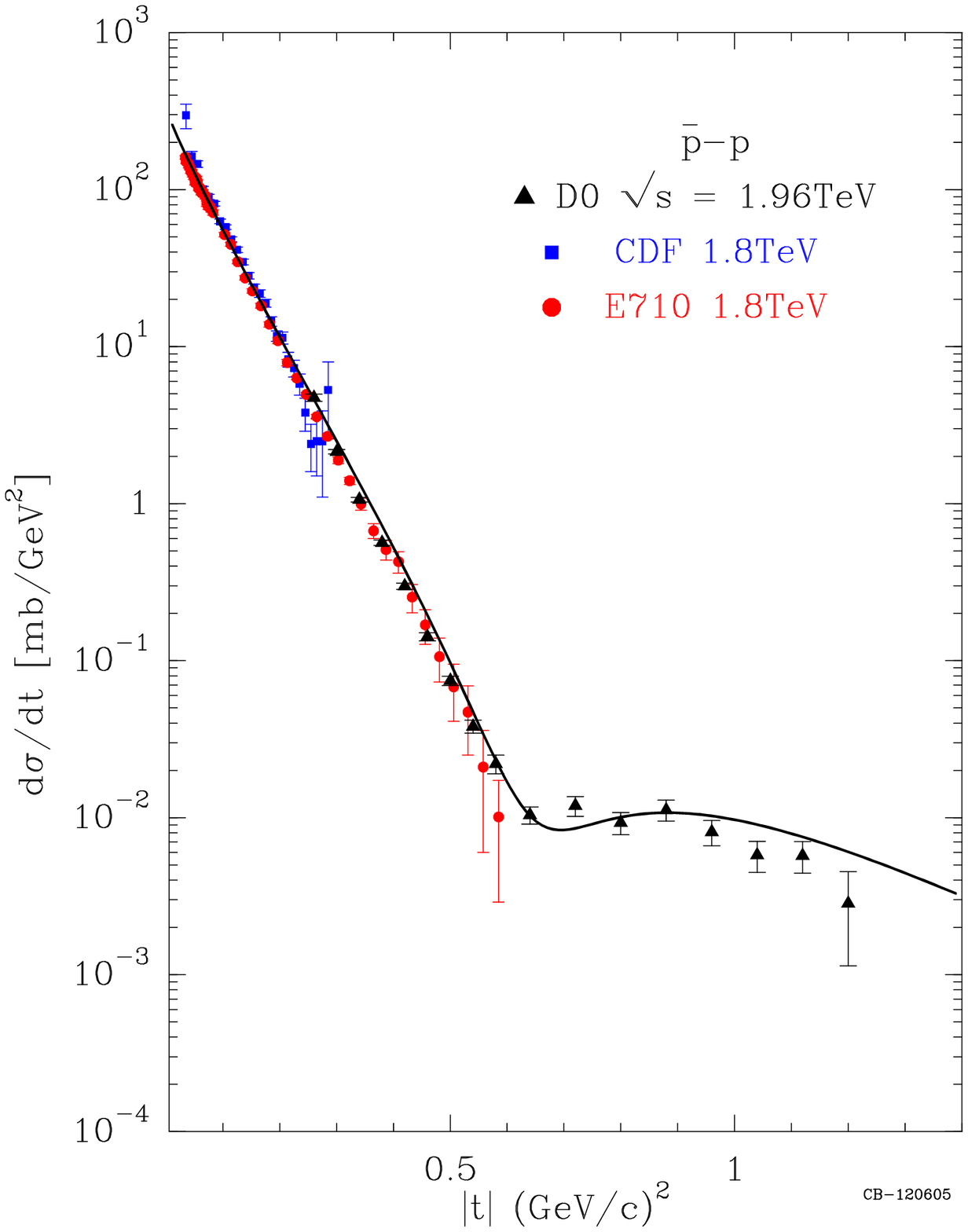,height=3.7in}
\end{center}
\vspace*{-10mm}
\caption{({\it Left}): The absolute value of the $\bar pp$ elastic 
scattering amplitude, {\it solid curve} $|\mbox{Im}~a(s,t)|$, 
{\it dashed curve} $|\mbox{Re}~a(s,t)|,$ versus $t$ at 
$\sqrt{s}=1.8\mbox{TeV}$. ({\it Right}): The corresponding $d\sigma/dt$, 
prediction from Ref. (10), new D0 data from Ref. (11).}
\label{fig:dsig-tevat}
\end{figure}

\begin{figure}[htb]
\begin{center}
\psfig{figure=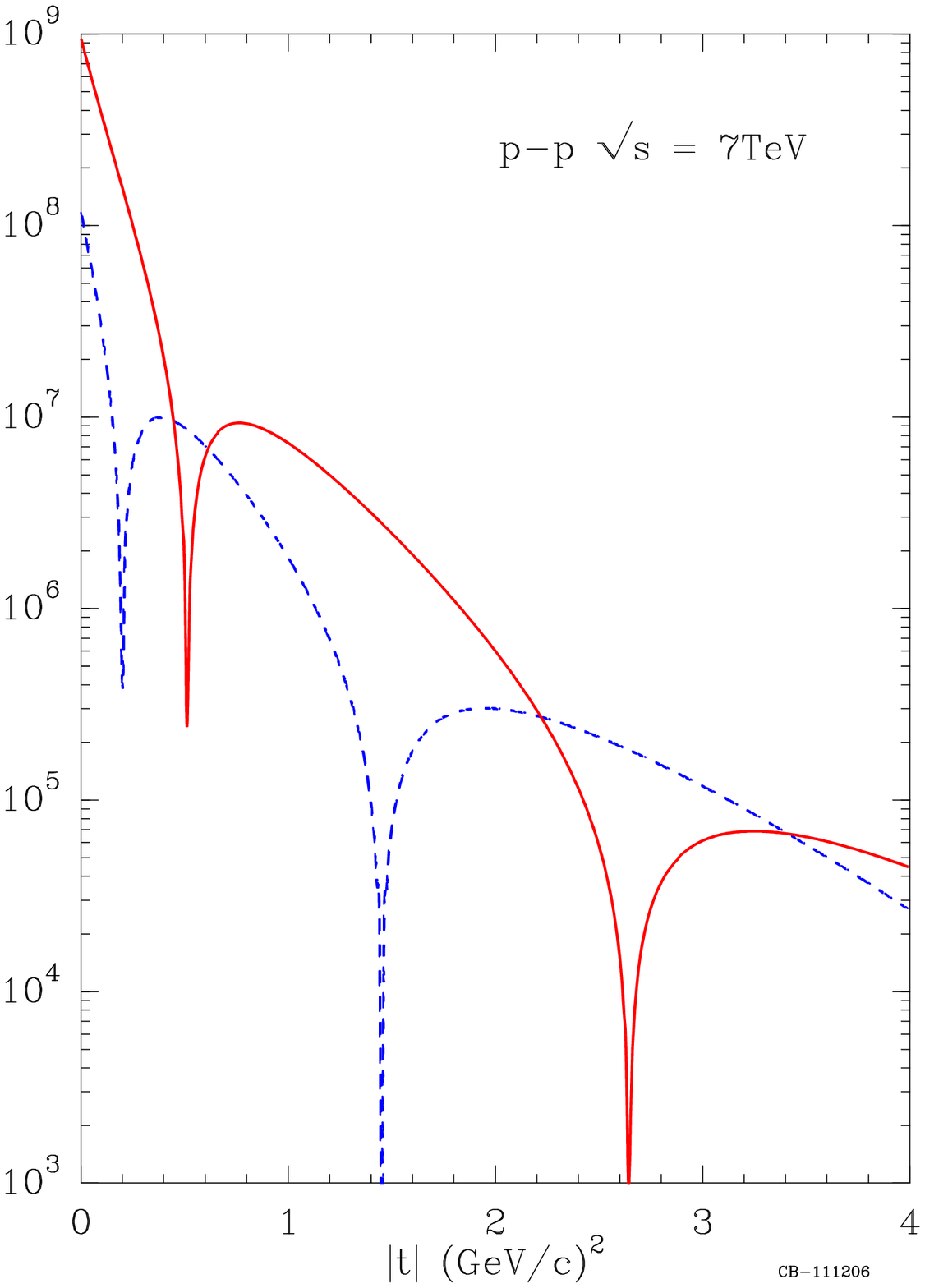,height=3.7in}
\psfig{figure=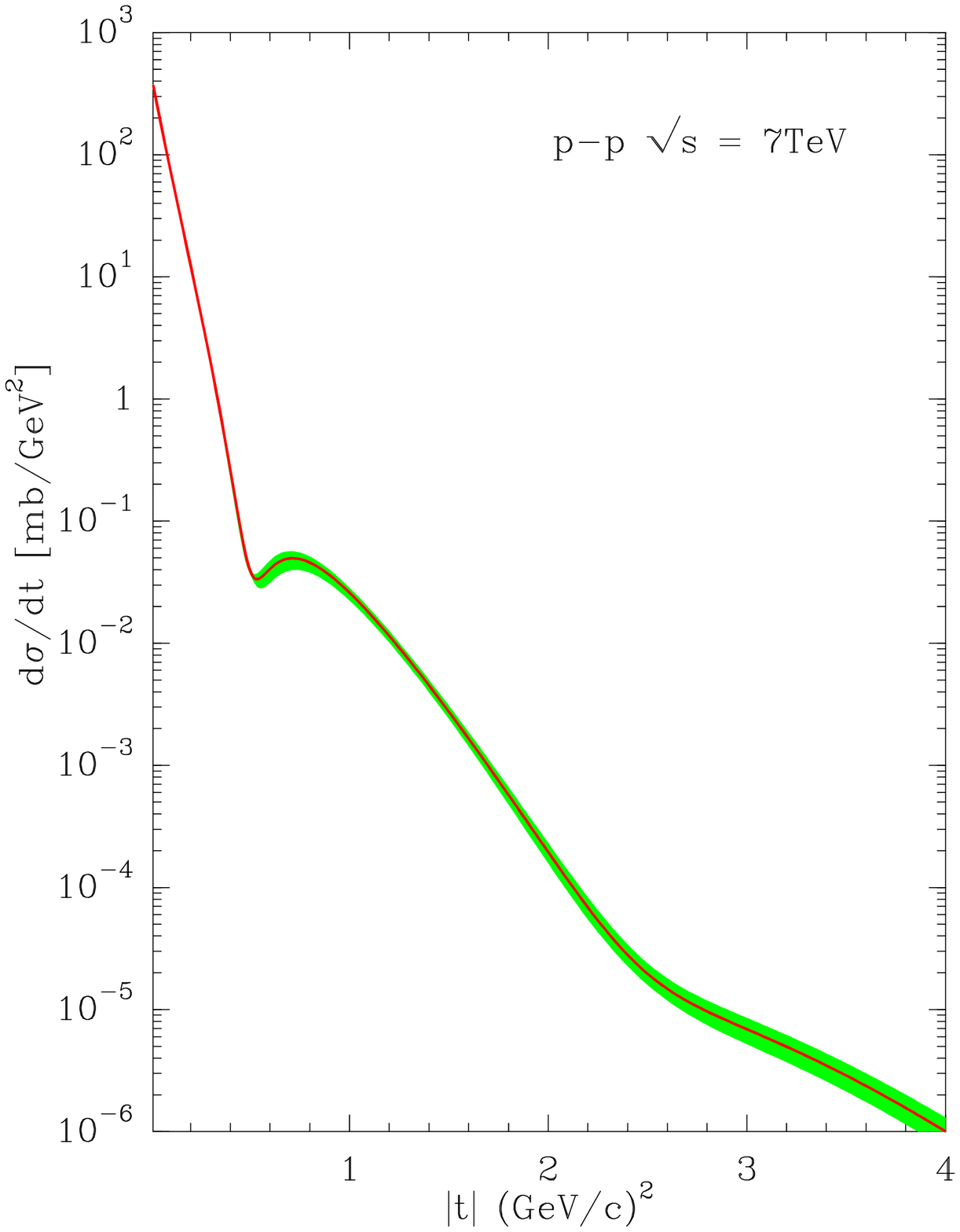,height=3.7in}
\end{center}
\vspace*{-10mm}
\caption{({\it Left}): The absolute value of the $pp$ elastic scattering amplitude, {\it solid curve} $|\mbox{Im}~a(s,t)|$, {\it dashed curve} $|\mbox{Re}~a(s,t)|,$ versus $t$ at $\sqrt{s}=7\mbox{TeV}$. ({\it Right}): The corresponding differential cross section (Taken from Ref.(12)).}
\label{fig:dsig-lhc}
\end{figure}

This kinematic region allows us to understand the behavior of the differential cross section from the $t$-dependence of the real and imaginary parts of the scattering amplitude, which have both some zeros at different $t$ values, as shown in Figs. 3 and 4 (Left). The imaginary part dominates over the real part, except when the imaginary part has a zero, producing either a shallow dip (or shoulder) for ${\bar pp}$ at $\sqrt{s}=1.8\mbox{GeV}$, as in Fig. \ref{fig:dsig-tevat} (Right) around $|t|=0.6 \mbox{GeV}^2$, or a real dip for $pp$ at $\sqrt{s}=7\mbox{GeV}$, as in Fig. \ref{fig:dsig-lhc} (Right) around $|t|=0.5 \mbox{GeV}^2$. Our prediction is in excellent agreement with the Tevatron data and although we predict the right position of the dip at LHC, we seem to underestimate the forward slope and to overestimate the cross section in the region of the second maximum, determined by TOTEM \cite{deile}.
\section{Concluding remarks}
LHC is opening up a new area for $pp$ elastic scattering and 
TOTEM has confirmed the following basic features expected at LHC from BSW:
 $\sigma_{tot}$ and $\sigma_{el}/\sigma_{tot}$ increase,
 the diffraction peak is still shrinking, the dip position is moving in and the second maximum is moving up.
 So far one observes only partial quantitative agreement with the BSW approach, but more data are needed, in particular from ATLAS-ALFA. One should not forget the relevance of the measurement of $\rho$.
\section*{Acknowledgments}
I am grateful to Prof. Chung-I Tan for organizing such an interesting scientific program. My warmest congratulations go to Prof. Tran Thanh Van for his wonderful project, in Quy Nhon, of the International Center for Interdisciplinary Science and Education (ICISE) in Vietnam, which will become soon a reality.

\end{document}